\documentclass[aps,prl, nofootinbib, twocolumn,superscriptaddress]{revtex4-2}
\usepackage{subfigure}
\usepackage{graphicx}
\usepackage{dcolumn}
\usepackage{bm}
\usepackage{xcolor}
\usepackage{float}
\definecolor{aa}{RGB}{0,0,139}

\newcommand{\DsToKKpi}{$D_s^\pm\to K^+K^-\pi^\pm$}

\newcommand{\Dss}{$D_s^{\ast\pm}$}

\newcommand{\eeToDssDss}{$e^+e^-\to D_s^{\ast+}D_s^{\ast-}$}
\newcommand{\pipiJpsi}{$\pi^+\pi^-J/\psi$}

\usepackage[colorlinks,urlcolor=aa,linkcolor=blue,anchorcolor=aa,citecolor=aa]{hyperref}
\usepackage[mathlines]{lineno}
\usepackage{multirow}
\usepackage{enumerate}
\usepackage{amsmath}
\usepackage{relsize}
\def\babar{\mbox{\slshape B\kern-0.1em{\smaller A}\kern-0.1em
    B\kern-0.1em{\smaller A\kern-0.2em R}}}

\begin{document}

\preprint{APS/123-QED}

\title{{\bf \boldmath Precise measurement of the $e^{+}e^{-}\rightarrow D_{s}^{\ast+}D_{s}^{\ast-}$ cross sections at center-of-mass energies from threshold to 4.95~GeV}}
\author{M.~Ablikim$^{1}$, M.~N.~Achasov$^{5,b}$, P.~Adlarson$^{75}$, X.~C.~Ai$^{81}$, R.~Aliberti$^{36}$, A.~Amoroso$^{74A,74C}$, M.~R.~An$^{40}$, Q.~An$^{71,58}$, Y.~Bai$^{57}$, O.~Bakina$^{37}$
% ...
\\
\vspace{0.2cm}
(BESIII Collaboration)\\
\vspace{0.2cm} {\it
$^{1}$ Institute of High Energy Physics, Beijing 100049, People's Republic of China\\
% ...
$^{5}$ Budker Institute of Nuclear Physics SB RAS (BINP), Novosibirsk 630090, Russia\\
% ...
$^{36}$ Johannes Gutenberg University of Mainz, Johann-Joachim-Becher-Weg 45, D-55099 Mainz, Germany\\
$^{37}$ Joint Institute for Nuclear Research, 141980 Dubna, Moscow region, Russia\\
% ...
$^{40}$ Liaoning Normal University, Dalian 116029, People's Republic of China\\
% ...
$^{57}$ Southeast University, Nanjing 211100, People's Republic of China\\
$^{58}$ State Key Laboratory of Particle Detection and Electronics, Beijing 100049, Hefei 230026, People's Republic of China\\
% ...
$^{71}$ University of Science and Technology of China, Hefei 230026, People's Republic of China\\
% ...
$^{74}$ University of Turin and INFN, (A)University of Turin, I-10125, Turin, Italy; (B)University of Eastern Piedmont, I-15121, Alessandria, Italy; (C)INFN, I-10125, Turin, Italy\\
$^{75}$ Uppsala University, Box 516, SE-75120 Uppsala, Sweden\\
% ...
$^{81}$ Zhengzhou University, Zhengzhou 450001, People's Republic of China\\
\vspace{0.2cm}
% ...
$^{b}$ Also at the Novosibirsk State University, Novosibirsk, 630090, Russia\\
% ...
}}

\vspace{0.4cm}
\date{\today}
\begin{abstract}
  The process $e^{+}e^{-}\rightarrow D_{s}^{\ast+}D_{s}^{\ast-}$ is studied with a semi-inclusive method using data samples at center-of-mass energies from threshold to 4.95~GeV collected with the BESIII detector operating at the Beijing Electron Positron Collider. The Born cross sections of the process are measured for the first time with high precision in this energy region. Two resonance structures are observed in the energy-dependent cross sections around 4.2 and 4.4~GeV. By fitting the cross sections with a coherent sum of three Breit-Wigner amplitudes and one phase-space amplitude, the two significant structures are assigned masses of (4186.5$\pm$9.0$\pm$30) MeV/$c^{2}$ and (4414.5$\pm$3.2$\pm$6.0)~MeV/$c^{2}$, widths of (55$\pm$17$\pm$53)~MeV and (122.6$\pm$7.0$\pm$8.2)~MeV, where the first errors are statistical and the second ones are systematic. The inclusion of a third Breit-Wigner amplitude is necessary to describe a structure around 4.79~GeV.

\end{abstract}

\pacs{Valid PACS appear here}
\maketitle

In $e^+e^-$ annihilations, several conventional vector
charmonium states are established in the inclusive
hadronic cross sections above the open charm threshold, such as the $\psi(3770)$,
$\psi(4040)$, $\psi(4160)$, and $\psi(4415)$. However, unexpected vector charmonium-like
resonance structures have also been discovered over the past two decades with a charmonium and light hadrons in the final state. These include $\psi(4260)$, initially observed in the $e^+e^-\to$\pipiJpsi~ process~\cite{ISRY4260ToPiPiJpsi-BaBar,
ISRY4260ToPiPiJpsi2007-Belle}, and $\psi(4360)$ and $\psi(4660)$, observed
in $e^+e^-\to \pi^+\pi^-\psi(3686)$~\cite{ISRPipiPsipBelle,ISRPipiPsipBabar}
by the $B$-factory experiments \babar\ and Belle.

The $\psi(4260)$ state has been searched for in several alternative decay modes, involving charmonium states
(such as $K\bar{K} J/\psi$~\cite{KKJpsiBelle2014,
KKJpsiBESIII2018}, $\eta J/\psi$~\cite{etaJpsiBelle2013,
etaJpsiBESIII2015}, $\eta^\prime J/\psi$~\cite{etapJpsiBESIII2016},
$\pi\pi h_c$~\cite{pipihcBESIII2013, pipihcBESIII2014},
$\pi^+\pi^-\psi(3686)$, and
$\omega\chi_{cJ}$~\cite{omegaChicJBESIII2015})
and charmed mesons (such as $D\bar{D}$~\cite{DDbar1,DDbar2}, $D^{\ast0}D^{\ast-}\pi^{+}+c.c.$~\cite{Dst0Dstpi} and $\bar{D}D^*\pi+c.c.$~\cite{DD0pi}).
A structure around 4.26~GeV is observed in most of the modes except in the $D\bar{D}$ mode~\cite{DDbar1,DDbar2}.

In 2017, a dedicated analysis of $e^+e^-\to \pi^+ \pi^- J/\psi$ performed by the BESIII experiment
revealed that the so-called $\psi(4260)$ state is actually a
combination of two states~\cite{pipiJpsiBESIII2017}. The main
component has a lower mass and a much narrower width than the initially measured state,
and is consistent with the structure observed in $e^+e^-\to
\pi^+\pi^- h_{c}$~\cite{pipihcBESIII2017} and $\omega\chi_{c0}$~\cite{omegaChicJBESIII2015}, denoted as $\psi(4230)$. Futhermore, the BESIII experiment also discovered another structure
 around $4.32~{\rm~GeV}/c^2$, with
a significance greater than $7.6\sigma$~\cite{pipiJpsiBESIII2017}. Interestingly, the mass of the $\psi(4230)$ state lies just at the production threshold of the $D_s^{\ast+}D_s^{\ast-}$ pair, which suggests a possible correlation between the
$\psi(4230)$ and this open-charm decay mode.

The measurements of the exclusive cross sections for charmed
strange meson pairs were performed by CLEO-c at center-of-mass energies ($E_\textrm{CM}$) up
to 4.26~GeV~\cite{openCharmXsec-CLEO}, and by
\babar~\cite{ISRDsXsec-BaBar} and Belle~\cite{ISRDsXsec-Belle} with the
initial-state radiation (ISR) method. Since these results are limited by either energy
range (CLEO-c) or statistics (\babar\ and Belle), it could not be concluded yet whether these vector charmonium-like states decay into charmed strange meson pairs or not.

With the data samples taken by the BESIII experiment, the cross sections
of \eeToDssDss~ can be measured from the production threshold up to 4.95~GeV, with much
improved precision over previous experiments. This measurement allows to investigate the
vector-chamonium(-like) structures, especially the $\psi(4230)$, therefore shedding light on their nature.

The BESIII detector is described in detail in~\cite{BESIIIdetector, detvis}.
The experimental data used in this analysis were taken at $E_\mathrm{CM}$
ranging from $4.226$~GeV (just above the $D_s^{\ast+}D_s^{\ast-}$ production threshold) to $4.95$~GeV
with 76 energy points~\cite{XYZEcm1,XYZEcm2,XYZEcm3} corresponding in total to an integrated luminosity of $15.67~\textrm{fb}^{-1}$~\cite{Lum1,Lum_Rscan,XYZEcm2}.
A {\scriptsize GEANT4}-based~\cite{geant4} Monte Carlo (MC) package, which includes the geometric description of the BESIII detector and its response, is used to produce simulated samples which are used to estimate the backgrounds and to determine the detection efficiencies and ISR corrections.

The \eeToDssDss~events are generated using helicity-amplitude (HELAMP) models in EvtGen~\cite{simulation1,simulation2} at all the energy points,
where the HELAMP inputs (relative magnitudes of the helicity amplitudes) are extracted from the helicity angular distributions of data, the width of \Dss~ is set to zero, and beam energy spread and ISR are considered
with the generator {\scriptsize \textsc{KKMC}}~\cite{kkmc1,kkmc2}. Possible background contributions are estimated with
inclusive MC samples generated by {\scriptsize \sc{KKMC}} with integrated luminosities comparable to data, in which the known decay modes are modeled with EvtGen using branching fractions taken from the
Particle Data Group~\cite{PDG}, and the remaining unknown charmonium decays
are modeled with {\sc lundcharm}~\cite{ref:lundcharm}. Final-state radiation~(FSR)
from charged particles is incorporated by the {\sc
photos} package~\cite{photos}.

To increase efficiency, a semi-inclusive method is performed by reconstructing only $D_s^{\ast+}$ or $D_s^{\ast-}$ of $e^+e^-\to D_s^{\ast+}D_s^{\ast-}$, with $D_s^{\ast\pm}\to\gamma D_s^{\pm}\to\gamma K^+K^-\pi^\pm$.
With this method at least one good photon, one pair of $K^+$$K^-$, and one $\pi^\pm$ are required in the final state.
The selection criteria for charged tracks and photon candidates are
described in Ref.~\cite{eventselection}.
To select $D_{s}^\pm$ candidates, the invariant mass $m_{KK\pi}$ of $K^+K^-\pi^\pm$ is required to be in the range $|m_{KK\pi}-m_{D_{s}}|<15$ MeV/$c^{2}$,
where $m_{D_{s}}$ is the nominal $D_{s}^\pm$ mass~\cite{PDG}.
All the $\gamma D_{s}^\pm$ (i.e. $\gamma K^+K^-\pi^\pm$) combinations are taken as $D_s^{\ast\pm}$ candidates for further selection.
The resolution of the missing mass for $D_s^{\ast\pm}$ candidates can be improved by
incorporating $M_\mathrm{miss}\equiv m_\mathrm{miss}+m_{\gamma KK\pi}-m_{D_s^\ast}$,
where $m_\mathrm{miss}$ and $m_{\gamma KK\pi}$ are the missing mass and the invariant mass of the $D_s^{\ast\pm}\to\gamma K^+K^-\pi^\pm$ candidates,  respectively,
and $m_{D_s^{\ast}}$ is the nominal mass of $D_s^{\ast\pm}$~\cite{PDG}.
With the signal MC events, the resolution of $M_\mathrm{miss}$ is estimated and parameterized
as an energy dependent function $\sigma^\mathrm{MC}_{M_\mathrm{miss}}(E_\mathrm{CM})$.
To select $D_s^{\ast\pm}$ from $e^+e^-\to D_s^{\ast+}D_s^{\ast-}$ events and suppress background,
the modified missing mass of $D_s^{\ast\pm}$ candidates is required to be in a window $|M_\mathrm{miss}-m_{D_s^\ast}|<5\sigma^\mathrm{MC}_{M_\mathrm{miss}}(E_\mathrm{CM})$.
To improve the mass resolution, $M_{\gamma KK\pi}\equiv m_{\gamma KK\pi}-m_{KK\pi}+m_{D_{s}}$ is calculated for $D_s^{\ast\pm}$ candidates.

\begin{figure}
 \centering
 \includegraphics[width=0.42\textwidth]{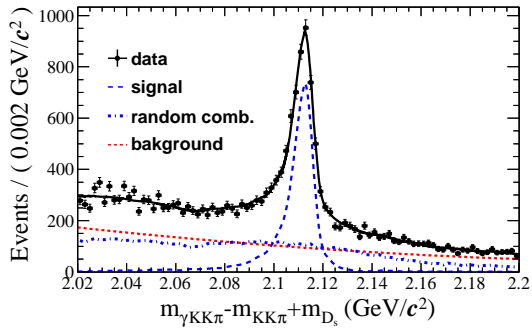}
 \caption{The $M_{\gamma KK\pi}$ distribution for data at 4.29 GeV. 
 The black curve  represents the fit, the red dashed curve the background, the blue dot-dashed curve the random combinations of $\gamma
 KK\pi$ from $e^+e^-\to D_s^{\ast\pm} D_s^{\ast\mp}$, and the blue dashed curve the correct $\gamma KK\pi$ combinations from the signal.}\label{pic:fitXYZData}
 \end{figure}

The yields of $D_{s}^{\ast\pm}$ signals are determined by fitting the $M_{\gamma KK\pi}$ distributions.
By matching the MC truth, correct $\gamma KK\pi$ combinations from \eeToDssDss~MC can be figured out, and the others from MC are called random combinations.
To describe the $D_{s}^{\ast\pm}$ signal in data, the $M_{\gamma KK\pi}$ shape of correct $\gamma KK\pi$ combinations from \eeToDssDss~MC is used. This MC signal shape is convolved with a Gaussian function to take into account the possible mass shift $\Delta m$, and the resolution (and $D_{s}^{\ast}$ width) difference $\Delta\sigma$ between data and MC simulation. The $M_{\gamma KK\pi}$ shape of random $\gamma KK\pi$ combinations from \eeToDssDss~MC is also used as one component and its ratio relative to the one for correct combinations is fixed according to the MC study.
To describe the background in $M_{\gamma KK\pi}$ distribution for data, a second order Chebyshev function $1+c_0 x+c_1(2x^2-1)$ with two coefficients $c_{0}$ and $c_{1}$ is used.
The parameters $\Delta m$, $\Delta \sigma$, $c_{0}$, and $c_{1}$ are floating initially in the fits. Since there is no $E_\mathrm{CM}$ dependence observed for $\Delta m$, $\Delta \sigma$, and $c_{1}$, they are finally fixed to the values averaged over $E_\mathrm{CM}$ in the fits to determine the nominal $D_{s}^{\ast\pm}$ signal yields, leaving only $c_0$ as floating.
As an example, the fit of $M_{\gamma KK\pi}$ to data at $E_\mathrm{CM}=4.29$ GeV \footnote{This is a nominal value and the measured one is $4287.88\pm0.06\pm0.34$ MeV~\cite{XYZEcm3}.} is shown in Fig.~\ref{pic:fitXYZData}.

A study of the inclusive MC samples shows that no peaking background is found from  events without $D_{s}^{\ast\pm}$ after applying all the selection criteria. The ISR-produced $D_{s}^{\pm}D_{s}^{\ast\mp}$ events can contribute as a peaking background, which is subtracted by normalized MC according to the cross sections of $e^{+}e^{-} \to D_s^{\pm}D_s^{\ast\mp}$ \footnote{A measurement by BESIII to be published in near future.} and the luminosities. The other two-body processes containing $D_{s}^{\ast\pm}$ are
$e^{+}e^{-}\to D_s^{\ast\pm}D_{sJ}^{(\ast)\mp}$, but they do not contribute to the peaking background since their missing mass (i.e. the mass of $D_{sJ}^{(\ast)\mp}$) is outside of the $D_{s}^{\ast\mp}$ mass region. For the same reason, the three-body process $e^{+}e^{-}\to D_{s}^{\ast}D^{(\ast)}K$ does not contribute as peaking backgrounds either. The contribution from the three-body process $e^{+}e^{-}\to D_{s}^{\ast\pm}D_{s}^{\mp}\pi^{0}$ is expected to be negligible due to isospin violation.

The Born cross sections $\sigma_\mathrm{Born}$ and the dressed cross sections $\sigma_\mathrm{dressed}$ at each energy point are calculated using
\begin{equation}\label{equ:xsection}
\begin{aligned}
    \sigma_\mathrm{Born}&=\sigma_\mathrm{dressed}|1+\Pi|^{2} \\
    &=\frac{ N_{D_s^\ast}^\mathrm{fit} - N_{D_{s}^{\pm}D_{s}^{\ast \mp}}}{2\mathcal{B}(D_s^\pm\to K^+K^- \pi^\pm) \epsilon(1+\delta)\frac{1}{|1+\Pi|^{2}}\mathcal{L}_\mathrm{int}},
\end{aligned}
\end{equation}
where $N_{D_s^\ast}^\mathrm{fit}$ is the fitted $D_s^\ast$ signal yield, $N_{D_{s}^{\pm}D_{s}^{\ast \mp}}$ is the number of the estimated peaking background events from the ISR produced $D_{s}^{\pm}D_{s}^{\ast\mp}$, $\mathcal{L}_\mathrm{int}$ is the integrated luminosity, $1+\delta$ is the ISR correction factor, $\frac{1}{|1+\Pi|^{2}}$ is the correction factor from the vacuum polarization (VP)~\cite{vpcorraction}, $\mathcal{B}(D_s^\pm\to K^+K^-\pi^\pm)$ is the branching fraction of the decay $D_s^\pm\to K^+K^-\pi^\pm$, which is taken from Ref.~\cite{PDG}, and  $\epsilon$ is the detection efficiency containing the branching fraction of $D_s^{\ast\pm}\to\gamma D_s^\pm$ \cite{PDG}. As $D_{s}^{\ast+}$ and $D_{s}^{\ast-}$ are not separated in the fitted signal yields, there is a factor of 2 in the denominator.

The ISR effect also impacts on the detection efficiency in addition to the correction factor. Since ISR depends on the dressed cross section line shape, an iterative procedure is used to determine the dressed cross sections~\cite{newiterationway}. First, signal MC samples are generated at all the energy points with a flat dressed cross section line shape using {\scriptsize \textsc{KKMC}} to determine the initial detection efficiencies, ISR correction factors and subsequently the preliminary measured dressed cross sections.
The measured dressed cross sections are fitted by a smooth line shape (defined later in Eq.~\eqref{Born_crosssection_lineshape}) as a function of $E_\textrm{CM}$.
With the fitted line shape, a MC-weighting method~\cite{newiterationway} is used to iteratively update the efficiencies, the ISR correction factors, the measured dressed cross sections, and the fitted line shape. After four iterations, the fit to the measured dressed cross section converged. By including the VP correction, the Born cross sections are summarized in the Supplemental Material~\cite{material}.

\begin{figure*}[!htbp]
 \centering
 \subfigure[Result 1]{
 \label{Fig.sub.1}
 \includegraphics[width=0.323\textwidth]{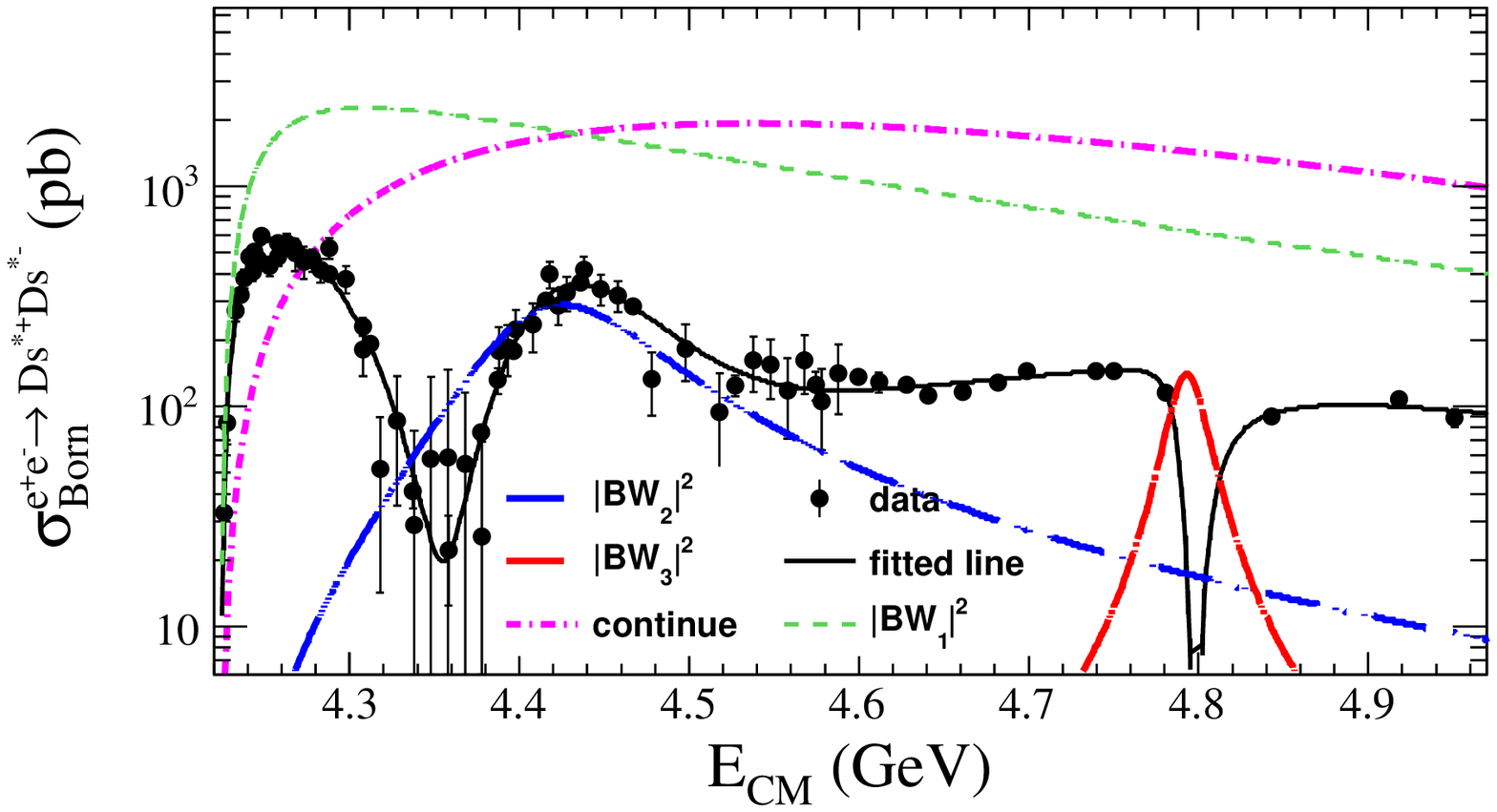}}
 \subfigure[Result 2]{
\label{Fig.sub.2}
 \includegraphics[width=0.323\textwidth]{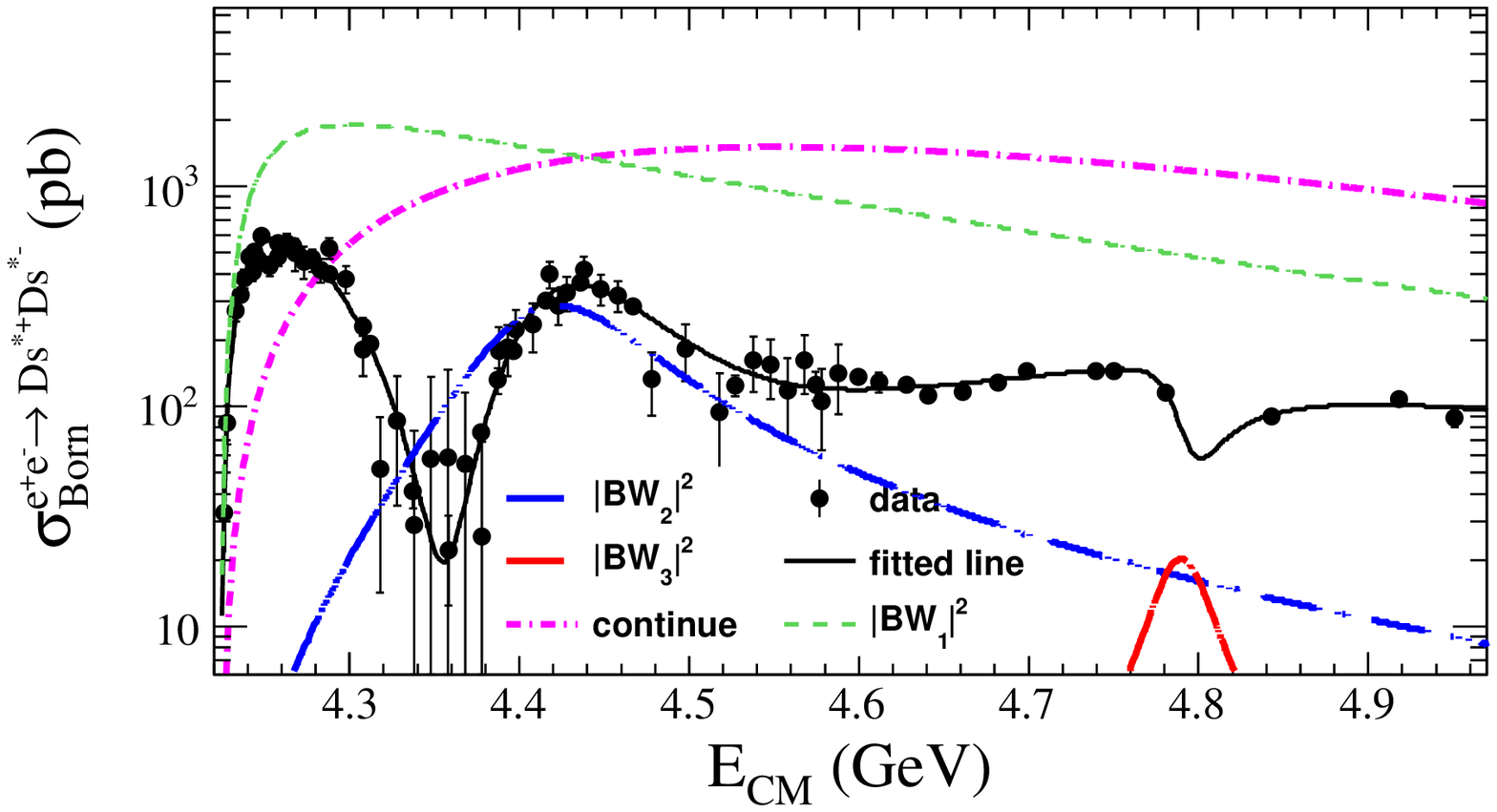}}
 \subfigure[Result 3]{
\label{Fig.sub.3}
 \includegraphics[width=0.323\textwidth]{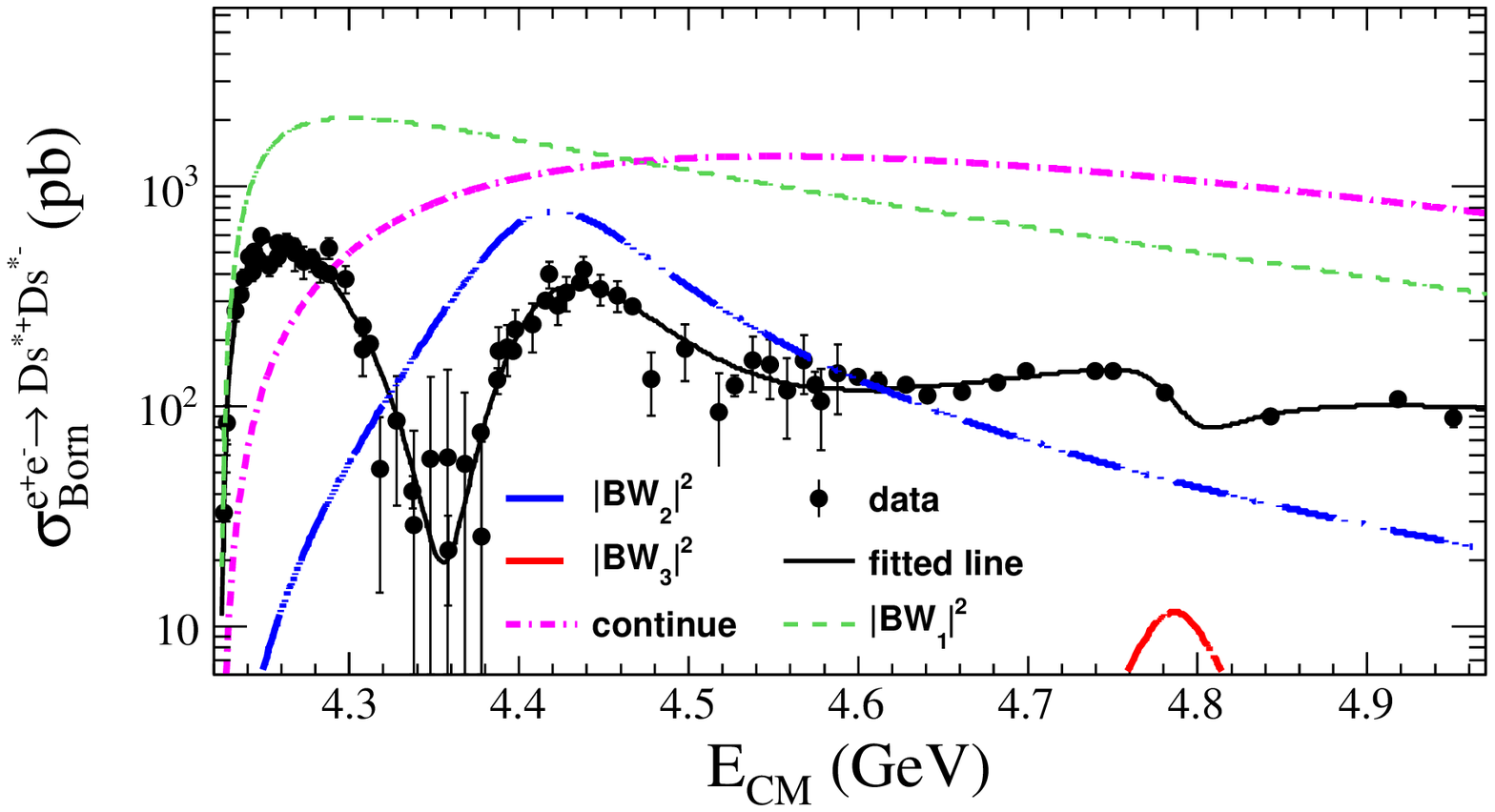}}
 \caption{
 Three fitting results for the measured Born cross sections of $e^{+}e^{-}\rightarrow D_{s}^{\ast+}D_{s}^{\ast-}$.
 The black dots with error bars are for the measured Born cross sections.
 In each plot, the black curve represents the fit; the green dashed, blue two-dashed and red long-dashed ones are for the three BWs, respectively,
 and the pink dot-dashed is for the PHSP contributions.
 }
  \label{pic:cs_fitting_results}
  \end{figure*}

The Born cross sections of $e^{+}e^{-}\to D_{s}^{\ast+}D_{s}^{\ast-}$ are fitted to estimate the parameters of the three structures
with a coherent sum of three Breit-Wigner (BW) and one two-body phase-space (PHSP) amplitudes assuming $D_{s}^{\ast+}D_{s}^{\ast-}$ in P-wave\footnote{$e^+e^-\to\gamma^\ast/Y\to D_{s}^{\ast+}D_{s}^{\ast-}$ is a vector$\to$vector+vector process, $P$-parity conservation requires the orbit angular momentum number $L$ of the $D_{s}^{\ast+}D_{s}^{\ast-}$ system
to be odd and $L=1$ (P-wave) is the most likely case.}

\begin{equation}\label{Born_crosssection_lineshape}
\begin{aligned}
\left |
\sum_{j=1}^{3}a_j\cdot e^{i\phi_{j}}\cdot BW_{j}(E_\mathrm{CM})
+\frac{a_{0}}{E_\mathrm{CM}^{n}}\right |^{2}\beta^{3}(E_{\mathrm{CM}}),
\end{aligned}
\end{equation}
where $BW_{j}(E_\mathrm{CM})=\frac{B_1(E_\mathrm{CM})}{(E^{2}_\mathrm{CM}-M_{j}^{2}+iM_{j}\Gamma_{j})\cdot{E_{\mathrm{CM}}}}$ is the $j$-th BW amplitude, the Blatt-Weisskopf function $B_1(E_\mathrm{CM})=\sqrt{1/(1+q^2(E_\mathrm{CM})R^2)}$ is used as the P-wave barrier factor~\cite{B_factor1, B_factor2} with $q(E_{\mathrm{CM}})$ the momentum of $D_{s}^{\ast\pm}$ and $R=1.6~\mathrm{GeV}^{-1}$~\cite{R1.6} the barrier radius,
$a_{j(0)}$ is the coefficient for the BW (the PHSP amplitude), $\beta(E_{\mathrm{CM}})$ is the velocity of $D_{s}^{\ast\pm}$,
and $\phi_{j}$ is the relative phase.
In the nominal fit of the Born cross sections, only statistical uncertainties are considered.
We find three sets of fitting results with comparable goodness of fit, which are shown in Fig.~\ref{pic:cs_fitting_results} and Table~\ref{tab:fittingresults}.
Due to the limited number of data points around 4.79~GeV, the fitted mass of the third structure
varies from 4786 to 4793.3 MeV/$c^{2}$ and the width from 27.1 to 60 MeV.
The statistical significance of the third structure exceeds 6.1$\sigma$ in all three fitting procedures, indicating that incorporating this amplitude leads to a more accurate description of the cross sections compared to using only two BWs. Besides the three BWs, one additional BW is tried to better describe the data points around 4.55~GeV, but it is not kept due to its small statistical significance.

  \begin{table}[!htbp]
  \centering 
  \caption{The fitting results of the Born cross sections.}\label{tab:fittingresults}
    \begin{tabular}{cccc}
    \hline
                                      & Result 1          & Result 2            & Result 3 \\
    \hline
  $M_1$ (MeV/$c^{2}$) & 4186.5$\pm$9.0      &4193.8$\pm$7.5        &4195.3$\pm$7.5\\
  $\Gamma_{1}$ (MeV) & 55$\pm$17              &61.2$\pm$9.0        &61.8$\pm$9.0\\
 $M_2$ (MeV/$c^{2}$) &  4414.5$\pm$3.2    &4412.8$\pm$3.2       &4411.0$\pm$3.2\\
   $\Gamma_{2}$ (MeV) &  122.6$\pm$7.0     & 120.3$\pm$7.0       &120.0$\pm$7.0\\
   $M_3$ (MeV/$c^{2}$) & 4793.3$\pm$7.5      &4789.8$\pm$9.0     &4786$\pm$10 \\
     $\Gamma_{3}$ (MeV) & 27.1$\pm$7.0    &41$\pm$39                &60$\pm$35 \\
    \hline
  \end{tabular}
  \end{table}

The systematic uncertainties for the measured Born cross sections are described as follows, some of which are common along all the energy points, while others are estimated depending on the $E_{\rm{CM}}$ range.

The systematic uncertainties of tracking (particle identification) efficiency are estimated to be 0.5\% (0.5\%) per $K^\pm$ and 0.2\% (0.4\%) per $\pi^\pm$ with a control sample of $D_s^\pm\to K^+K^-\pi^\pm$ decays~\cite{BrDsToKKpi}; thus a 1.2\% (1.4\%) systematic uncertainty is assigned for the tracking efficiency (particle identification)
in the $D_s^\pm\to K^+K^-\pi^\pm$ candidates selection.
The systematic uncertainty in the efficiency for photon reconstruction is set conservatively to 1\% based on a study with a sample of
$J/\psi\to\rho\pi$ events~\cite{BrChicToPi0pi0AndEtaEta-BES3}.
From fits of the invariant mass spectrum of $D_s^\pm\to K^+K^-\pi^\pm$ candidates and of the modified missing mass of $D_{s}^{\ast\pm}\to\gamma K^+K^-\pi^\pm$ candidates, the efficiencies for signal in the mass window for both data and MC samples can be calculated, and the relative differences in efficiency between data and MC simulation are taken as systematic uncertainties for the $D_s^\pm$ mass window and the modified missing mass window, respectively, which cover the possible resolution difference and the zero width setting for $D_{s}^{\ast\pm}$ in MC simulation.
The maximum difference in the dressed cross sections between the last two iterations, $0.2\%$, is taken as the systematic uncertainty for the stability of the iteration results. 
The systematic uncertainty of the branching fraction for \DsToKKpi~and $D_s^{\ast\pm}\to \gamma D_s^\pm$ is taken from Ref.~\cite{PDG}.
The integrated luminosities are measured by QED events~\cite{XYZlumi} with a systematic uncertainty of 1\%.
The uncertainty from VP correction is 0.1\%~\cite{vpcorraction}.
The uncertainty of the peaking background subtraction from the $e^+e^-\to D_{s}^\pm D_{s}^{\ast\mp}$ process is estimated to be $1\%$ mainly from the uncertainties of cross section measurement.
Instead of the HELAMP model, the PHSP model~\cite{simulation1,simulation2} is also used to generate $e^{+}e^{-}\to D_{s}^{\ast\pm}D_{s}^{\ast\mp}$ events
and the maximum difference in efficiency, 2.2\%, is taken as the systematic uncertainty for the generation model.
We get the total common systematic uncertainty by adding them in quadrature, which is 4.0\%. 

\begin{table*}[!htbp]
  \centering
  \caption{Relative systematic uncertainties of the measured Born cross sections.
  As the effects and the data statistics are energy dependent, some systematic uncertainties are estimated in several energy intervals
combining some data sets with low luminosities.
  }\label{tab:sysSum2}
   \begin{tabular}{c|c|c|c|c|c|c|c|c|c}
    \hline
    \multicolumn{1}{c|}{Sources} & \multicolumn{9}{c}{Systematic uncertainties (\%)} \\
    \hline
    \multicolumn{1}{c|} {Common} & \multicolumn{9}{c}{4.0} \\
    \hline
     \multicolumn{1}{c|}{Energy (interval) (GeV)}& 4.226 &4.228  &4.233&  4.233$\sim$4.24 & 4.24$\sim$4.3 & 4.3$\sim$4.4 & 4.4$\sim$4.82 &4.843  &4.86$\sim$4.95\\
    \hline
    Signal yields fitting with fixed parameters           &\multicolumn{5}{c|}{2.0}   &5.0  &\multicolumn{3}{c}{2.0}  \\
    \cline{1-10}
    Signal yields fitting range                           &\multicolumn{5}{c|}{4.0}   &5.0  &\multicolumn{3}{c}{4.0}  \\
    \cline{1-10}
     Cross section line shape description   &\multicolumn{4}{c|}{5.0}   &2.0  &5.0  &2.0  &5.0   &2.0  \\
    \cline{1-10}
     $E_\textrm{CM}$ uncertainty &24.5 &20.0  &9.4  &\multicolumn{6}{c}{0.8}               \\
    \hline
    \multicolumn{1}{c|}{Total}    &25.7 &21.5  &12.2 &7.9  &6.4  &9.6  &6.4  &7.9   &6.4  \\
    \hline
  \end{tabular}
\end{table*}

\begin{table} [!htbp]
  \centering 
  \caption{The systematic uncertainties of the fitted masses and widths. The systematic uncertainties of the fitted masses and widths. Fitting represent the uncertainty from the multiple fitting results. $R$ represent the uncertainty from the barrier radius, $E_\textrm{CM}$ the systematic errors from the $E_\mathrm{CM}$ uncertainty and the $\sigma_\textrm{Born}$ the systematic uncertainties on the measured Born cross sections.}\label{tab:fitting_cs_systematic}
    \begin{tabular}{ccccccc}
    \hline
    Sources&              Fitting&    $R$&   $E_\textrm{CM}$&    $\sigma_\textrm{Born}$&   Total \\
    \hline
    $M_{1}$ (MeV/$c^{2}$)&      8.8&    2.6&    28.6&      4.8&      30           \\
    $\Gamma_{1}$ (MeV)&         6.8&    1.9&    51&        11.8&     53        \\
    $M_{2}$ (MeV/$c^{2}$)&      3.5&    0.7&    3.9&       2.9&      6.0      \\
    $\Gamma_{2}$ (MeV)&         2.6&    0.3&    7.7&       0.9&      8.2     \\
    $M_{3}$ (MeV/$c^{2}$)&      7.3&    1.0&    2.4&       5.1&      9.3      \\
    $\Gamma_{3}$ (MeV)&         32.9&   1.1&    5.3&       3.4&      34       \\
    \hline
    \end{tabular}
\end{table}

The shape related parameters $\Delta m$, $\Delta\sigma$, and $c_1$ are fixed to the averaged values in the nominal fit (see Fig.~\ref{pic:fitXYZData} for an example).
The differences of the fitted signal yields, when these parameters are floating, are within
statistical uncertainties. However, to cover these differences conservatively, the whole energy range is divided into three intervals $(4.226, 4.3)$ GeV, $(4.3, 4.4)$ GeV, and $(4.4, 4.95)$ GeV
with assigned systematic uncertainties 2\%, 5\%, and 2\%, respectively, due to the signal and background shapes.
The boundaries of the nominal fitting range for $M_{\gamma KK\pi}$, which is $[2.02,~2.20]$~GeV$/c^2$, are changed by $10$ MeV to estimate  the corresponding fitting range uncertainties.
These are assigned to be 4\%, 5\%, and 4\%, respectively,
in the energy intervals $(4.226, 4.3)$ GeV, $(4.3, 4.4)$ GeV, and $(4.4, 4.95)$ GeV.
Eq.~\eqref{Born_crosssection_lineshape}  is used to fit the data iteratively
and get the converged dressed cross sections, during that the cross section line shape is similar to the one in Fig.~\ref{Fig.sub.1}.
Other two different line shapes with comparable fitting goodness, that are similar to the results shown in Fig.~\ref{Fig.sub.2} and \ref{Fig.sub.3},
plus an additional line shape obtained by the LOWESS (LOcally WEighted Scatterplot Smoothing)~\cite{Lowess1,Lowess2}, are used to calculate the systematic
uncertainties for the line shape description by repeating the iterations and taking the
differences in the results.
The systematic uncertainty of the measured $E_\mathrm{CM}$ is found to be less than 0.8 MeV (0.6 MeV) for $4.226<E_\mathrm{CM}<4.6$~GeV ($4.6<E_\mathrm{CM}<4.95$~GeV)~\cite{XYZEcm1,XYZEcm2,XYZEcm3} and it is used to shift all the energy points to conservatively estimate the impacts on the measured cross sections.
Since the cross section line shape varies dramatically near the threshold, the $E_\mathrm{CM}$ uncertainty could have significant impact on the ISR correction factors nearby,
subsequently affecting the measured cross sections and the fitting results around the threshold. The MC samples at the first four $E_\mathrm{CM}$ points near the threshold are regenerated with the shifted energies to update ISR corrections. After the iterations with the shifted energy points and the updated ISR correction factors near threshold, the resulting differences in the Born cross sections are taken as the systematic uncertainties due to the $E_\mathrm{CM}$ uncertainty.
These systematic uncertainties are summarized in Table~\ref{tab:sysSum2} by assuming no correlation among different sources.

The sum of three BW amplitudes with the one for the phase-space is an imperfect model, but it is useful to describe the cross section line shape smoothly for the
iterative ISR correction procedure
and to obtain masses and widths of the resonance(-like) structures for reference.
The systematic uncertainties of the fitted masses and widths of the three resonances are listed in Table~\ref{tab:fitting_cs_systematic}.
There are three fitting results as listed in Table~\ref{tab:fittingresults} with comparable goodness of fit.
The fitting result 1 is taken as the nominal one, and the biggest differences in the fitted central values between result 1 and the other two
are taken as the systematic uncertainties from the multiple fitting results.
The parameter $R$ is changed from 1.6 in the nominal fit to 5 GeV$^{-1}$ (corresponds the scale of the strong interaction which is about 1 fm)~\cite{R1.6,R5.0}, and the variations in the fitting results are taken as the corresponding systematic uncertainties.
The uncertainty from the measured $E_\mathrm{CM}$ has impact not only on the cross sections but also on the fitting results.
After the iterations with the shifted energies and the updated ISR correction factors,
the differences in the fitting results are considered as the uncertainty from the $E_\mathrm{CM}$ uncertainty.
The systematic uncertainties on the measured Born cross sections consist of a common part and an uncommon part.
The common part is the same for all energies and has no effect on fitted masses and widths of the BW functions.
The uncommon part is included to redo the cross section fitting and
the resulting differences are taken as systematic uncertainties.

In summary, with the world's largest $e^+e^-$ scan data sample between $4.226$
and $4.95$~GeV accumulated by BESIII, the Born cross sections of
\eeToDssDss~ are measured precisely.
Two enhancements in the $E_{\rm CM}$ dependent cross sections are observed around 4.2 and 4.45~GeV.
The $E_{\rm CM}$ dependent cross section line shape is modeled with a sum of three BW and one PHSP amplitudes.
The fitted mass and width for the first resonance are (4186.5$\pm$9.0$\pm$30) MeV/$c^{2}$ and (55$\pm$17$\pm$53)~MeV, respectively.
These results are consistent with the $\psi(4160)$ observed in the inclusive cross section of $e^+e^-\to \textrm{hadrons}$~\cite{R_bes2} and in the dimuon spectrum of $B^-\to K^-\mu^+\mu^-$~\cite{B2Kmumu2013}.
While considering the systematic uncertainties, these results are also consistent with $\psi(4230)$ observed in the $\pi^+\pi^-J/\psi$ mode. If the contribution is from $\psi(4230)$, it indicates that the $\psi(4230)$ couples more strongly to the $D_s^{\ast+}D_s^{\ast-}$ mode than to the modes with charmonium states, since the cross section of $e^+e^-\to D_s^{\ast+}D_s^{\ast-}$ at 4.23~GeV is roughly one order of magnitude higher than that of $e^+e^-\to \pi^+\pi^-J/\psi$.
This information is vital to understand the nature of $\psi(4230)$,
especially considering the fact that its mass is very close to
the production threshold of $D_s^{\ast+}D_s^{\ast-}$.  The fitted mass and width for the second resonance are (4414.5$\pm$3.2$\pm$6.0) MeV/$c^{2}$ and (122.6$\pm$7.0$\pm$8.2) MeV, respectively.
The mass is consistent with $\psi(4415)$, and the measured width from this work is a bit higher than the world averaged value of $\psi(4415)$~\cite{PDG}, but still within three standard deviations. If we assume that this resonance is the $\psi(4415)$, this is the first time that this state is observed in the $D_s^{\ast+}D_s^{\ast-}$ decay mode. An additional third BW amplitude describes the $E_{\rm CM}$ dependent cross sections around 4.79~GeV better than just using two BW functions, with statistical significance greater than $6.1\sigma$, however,
more data points in the vicinity are needed for further clarification~\cite{whitepaper}.
It is out of the scope of this paper, but a unitary approach based on $K$-matrix formalism to fit the cross-section results of various exclusive channels is expected to be carried out as a more comprehensive and robust analysis of the vector-charmonium(-like) structures (for instance, an analysis for vector bottomonia with this method can be found in Ref.~\cite{K-matrix}).

\begin{acknowledgments}
The BESIII Collaboration thanks the staff of BEPCII and the IHEP computing center for their strong support. This work is supported in part by National Key R\&D Program of China under Contracts Nos. 2020YFA0406300, 2020YFA0406400; National Natural Science Foundation of China (NSFC) under Contracts Nos. 12150004, 11635010, 11735014, 11835012, 11935015, 11935016, 11935018, 11961141012, 12022510, 12025502, 12035009, 12035013, 12061131003, 12192260, 12192261, 12192262, 12192263, 12192264, 12192265, 12221005, 12225509, 12235017, U1732105, U1632106; Program of Science and Technology Development Plan of Jilin Province of China under Contract No. 20210508047RQ; The Fundamental Research Funds of Shandong University; the Chinese Academy of Sciences (CAS) Large-Scale Scientific Facility Program; the CAS Center for Excellence in Particle Physics (CCEPP); CAS Key Research Program of Frontier Sciences under Contracts Nos. QYZDJ-SSW-SLH003, QYZDJ-SSW-SLH040; 100 Talents Program of CAS; The Institute of Nuclear and Particle Physics (INPAC) and Shanghai Key Laboratory for Particle Physics and Cosmology; ERC under Contract No. 758462; European Union's Horizon 2020 research and innovation programme under Marie Sklodowska-Curie grant agreement under Contract No. 894790; German Research Foundation DFG under Contracts Nos. 443159800, 455635585, Collaborative Research Center CRC 1044, FOR5327, GRK 2149; Istituto Nazionale di Fisica Nucleare, Italy; Ministry of Development of Turkey under Contract No. DPT2006K-120470; National Research Foundation of Korea under Contract No. NRF-2022R1A2C1092335; National Science and Technology fund of Mongolia; National Science Research and Innovation Fund (NSRF) via the Program Management Unit for Human Resources \& Institutional Development, Research and Innovation of Thailand under Contract No. B16F640076; Polish National Science Centre under Contract No. 2019/35/O/ST2/02907; The Swedish Research Council; U. S. Department of Energy under Contract No. DE-FG02-05ER41374.
\end{acknowledgments}

\end{document}